\documentclass[twocolumn,nofootinbib,prd,aps,superscriptaddress,tightenlines]{revtex4}

\usepackage{amssymb}

\usepackage{slashed}
\usepackage{graphicx}

{\count255=\time\divide\count255 by 60 \xdef\hourmin{\number\count255}
  \multiply\count255 by-60\advance\count255 by\time
  \xdef\hourmin{\hourmin:\ifnum\count255<10 0\fi\the\count255}}

\newcommand{\nn}{\nonumber \\ }
\newcommand\eUV{\epsilon_{\text{UV}}}

\newcommand\lQ{\mathsf{L_Q}}
\newcommand\lM{\mathsf{L_M}}

\def\bn{ \bar n}

\newcommand{\beq}{\begin{equation}}
\newcommand{\eeq}{\end{equation}}
\newcommand{\bea}{\begin{eqnarray}}
\newcommand{\eea}{\end{eqnarray}}
\newcommand{\co}{\, ,}
\newcommand{\fs}{\, .}
\newcommand{\al}{\alpha}

\def\scetew{ $\text{SCET}_{\text{EW}}$}

\begin{document}

\title{Soft-Collinear Factorization and Zero-Bin Subtractions}

\author{Jui-yu Chiu}
\affiliation{Department of Physics, University of California at San Diego,
  La Jolla, CA 92093}

\author{Andreas Fuhrer}
\affiliation{Department of Physics, University of California at San Diego,
  La Jolla, CA 92093}

\author{Andr\'e~H.~Hoang}
\affiliation{Max-Planck-Institut f\"ur Physik (Werner-Heisenberg-Institut), 
F\"ohringer Ring 6, 80805 M\"unchen, Germany}

\author{Randall Kelley}
\affiliation{Department of Physics, University of California at San Diego,
  La Jolla, CA 92093}

\author{Aneesh V.~Manohar}
\affiliation{Department of Physics, University of California at San Diego,
  La Jolla, CA 92093}

\begin{abstract}

We study the Sudakov form factor for a spontaneously broken gauge theory using
a (new) $\Delta$-regulator. To be well-defined, the effective theory requires
zero-bin subtractions for the collinear sectors. The zero-bin subtractions
depend on the gauge boson mass $M$ and are  not scaleless. They have both
finite and $1/\epsilon$ contributions, and are needed  
to give the correct anomalous dimension and low-scale matching
contributions. We also demonstrate the necessity of zero-bin subtractions for 
soft-collinear factorization. We find that after zero-bin
subtractions the form factor is the sum of the collinear contributions
\emph{minus} a soft mass-mode contribution, in agreement with a previous
result of Idilbi and Mehen in QCD. This appears to conflict with
the method-of-regions approach, where one gets the 
sum of contributions from different regions. 
\end{abstract}

\date{\today\quad\hourmin}

\maketitle

\section{Introduction}

Soft collinear effective theory (SCET) \cite{BFL,SCET} is a field theory which
describes the interactions of energetic particles with small invariant
mass. SCET was originally developed for QCD processes, but has
recently~\cite{CGKM1} been extended to broken gauge theories with massive
gauge bosons. This allows one to compute electroweak corrections to Standard
Model processes at high energies, and to sum electroweak Sudakov
logarithms~\cite{CGKM2,CKM}.

Applications of SCET to electroweak processes require evaluating collinear and
soft Feynman graphs with massive gauge bosons. These graphs are not-well
defined, even in dimensional regularization with an off-shellness, and require
additional regularization. In Refs.~\cite{CGKM1,CGKM2,CKM}, the graphs were
evaluated with an analytic regulator~\cite{bef,analytic}; the individual
diagrams depend on the analytic regulator parameters, but the total amplitude
is regulator independent. The analytic regulator has some unpleasant
properties with regards to gauge invariance and factorization, two essential
ingredients of SCET. We propose a convenient new regulator, called the
$\Delta$-regulator, which can be implemented directly on the level of the SCET
Lagrangian without the need to define further prescriptions for computing
diagrams. This regulator is similar to using a mass, and unlike off-shellness,
it regulates diagrams with massive gauge bosons. 

The $\Delta$-regulator is used to compute the Sudakov form factor using SCET
for a spontaneously broken $SU(2)$ gauge theory with a common gauge boson mass
$M$. This form factor was computed
previously in Ref.~\cite{CGKM1} using an analytic regulator. We discuss the
factorization structure of the effective theory using the
$\Delta$-regulator. As noted previously~\cite{collins,leesterman} in the
framework of QCD, the
amplitudes only factorize when the collinear 
sectors are defined including zero-bin subtractions~\cite{MS}, to avoid double-counting 
the soft momentum region. We show in the broken $SU(2)$ gauge theory that the
usual construction of collinear gauge interactions into 
collinear Wilson lines, while true at tree level, is valid at the loop-level
only when the collinear sector is defined with zero-bin subtractions.

Recently Idilbi and Mehen~\cite{IM1,IM2} reemphasized the
necessity for zero-bin subtractions~\cite{MS}. They studied deep inelastic scattering
and showed that the correct total amplitude has the form $I_n+I_{\bn}-I_{s}$,
the sum of the $n$-collinear, $\bn$-collinear, {\it minus} the soft
contributions, rather than the naive expectation 
$I_n+I_{\bn}+I_{s}$. The sign change of $I_s$ arises because the collinear
contributions have to be properly thought of as zero-bin subtracted, $I_n \to
I_n-I_s$. This converts the second (incorrect) form of the result into the
first. In the case of deep inelastic scattering the effective
theory graphs are scaleless, and so vanish in dimensional regularization. The
net effect of the zero-bin subtractions is therefore to simply convert
$1/\epsilon$ infrared divergent poles into $1/\epsilon$ ultraviolet divergent
poles. This might lead one to think of the zero-bin subtraction as an academic
issue. However, the conversion is necessary to obtain the correct form of the
anomalous dimensions. In the case 
of a broken $SU(2)$ gauge theory we also find that the zero-bin subtractions
are mandatory. In contrast to deep inelastic scattering, the effective theory
graphs  depend on the gauge boson mass $M$, and are no longer scaleless. As a
result, the zero-bin subtractions are necessary not just to convert infrared
divergences into ultraviolet ones, but also to correctly obtain the finite
parts of the diagram. 

The article is organized as follows: We start out with a discussion of the
full theory and the SCET formalism in section \ref{sec:ft}, and in section
\ref{sec:wilson}, we discuss how Wilson line regularization breaks
factorization. The $\Delta$-regulator is introduced in
section~\ref{sec:delta}. The effective theory computation, zero-bin
subtractions, gauge dependence and 
momentum regions  are discussed in section~\ref{sec:calc}. The conclusions are
given in section \ref{sec:sum}. Some technical details are relegated to
appendix~\ref{app:im}. 

\section{Formalism}\label{sec:ft}

The theory we consider is a $SU(2)$ spontaneously broken gauge theory, with a
Higgs in the fundamental representation, where all gauge bosons have a common
mass, $M$. This is the theory used in many previous
computations~\cite{kps,fkps,jkps,jkps4,js}, and allows us to compare with
previous results.   It is convenient, as in Ref.~\cite{js}, to write the group
theory factors using $C_F$, $C_A$, $T_F$\footnote{Note that the results only
  hold for $C_A=2$, since for an SU$(N)$ group with $N>2$, a fundamental Higgs
  does not break the gauge symmetry completely.}  

The physical quantity of interest is the Sudakov form factor $F(Q^2)$ in the
Euclidean region,  
\beq\label{eq:sff}
F(Q^2) = \langle p_2 | \bar{\psi} \Gamma \psi | p_1 \rangle \co
\eeq
where $Q^2 = -(p_2-p_1)^2 \gg M^2$ and $\Gamma$ is a generic Dirac
structure. In SCET, $F(Q^2)$ is computed in three steps: (i) matching from the
full gauge theory to SCET at $\mu=Q$ (high-scale matching) (ii) running in
SCET between $Q$ and $M$ and (iii) integrating out the gauge bosons at $\mu=M$
(low-scale matching). The high-scale matching computation is given in
Ref.~\cite{CGKM1}. The SCET computation of the running and low-scale matching
is discussed in this article. All computations are done to leading order in
SCET power counting, i.e.\ neglecting $M^2/Q^2$ power corrections. 

\begin{figure}
\begin{center}
\includegraphics[width=6cm]{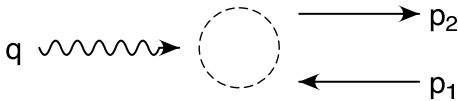}
\end{center}
\caption{Breit frame kinematics. \label{fig:back}}
\end{figure}

The SCET fields and Lagrangian depend on two null four-vectors $n$ and $\bar
n$,  with $n=(1,\bf{n})$ and $\bn=(1,-\bf{n})$, where $\bf{n}$ is a unit
vector, so that $\bar n \cdot n=2$. In the Sudakov problem, one works in the
Breit frame, with $n$ chosen to be along the $p_2$ direction, so that $\bar n$
is along the $p_1$ direction. In the Breit frame, the momentum transfer $q$
has no time component, $q^0=0$, so that the particle is back-scattered (see
Fig.~\ref{fig:back}).  
The light-cone components of a four-vector $p$ are defined by $p^+ \equiv n
\cdot p$, $p^- \equiv \bn \cdot p$, and $p_\perp$, which is orthogonal to $n$
and $\bn$, so that 
\begin{eqnarray}
p^\mu &=& \frac 12 n^\mu (\bn \cdot p)+\frac12 \bn^\mu (n \cdot p) + p_\perp^\mu.
\label{3}
\end{eqnarray}
In our problem, $p_1^-=p_{1\perp}=p_2^+=p_{2\perp}=0$, and $Q^2=p_1^+ p_2^-$.
A fermion moving in a direction close to $n$ is described by the $n$-collinear
SCET field $\xi_{n,p}(x)$, where $p$ is a label momentum, and has components
$\bar n \cdot p$ and $p_\perp$~\cite{BFL,SCET}.  It describes particles (on-
or off-shell) with energy $2E\sim\bar n \cdot p$, and $p^2 \ll Q^2$. The total momentum of the field $\xi_{n,p}(x)$ is $p+k$, where $k$ is the residual momentum of order $Q \lambda^2$ contained in the Fourier
transform of $x$. Note that the label momentum $p$ only contributes to the
minus and $\perp$ components of the total momentum.  

The massive gauge fields are represented by several distinct fields in the
effective theory: $n$-collinear fields $A_{n,p}(x)$ and $\bar n$-collinear fields
$A_{\bar n,p}(x)$ with labels, and so-called mass-mode fields
$A_s(x)$~\cite{Fleming:2007xt,Fleming:2007qr} to which 
we do not give any label. This is analogous to the label conventions for soft
and ultrasoft fields introduced in NRQCD~\cite{lmr}. The 
$n$-collinear field contains massive gauge bosons with momentum near the
$n$-direction, and momentum scaling $\bar n \cdot p \sim Q$, $n \cdot p \sim Q
\lambda^2$, $p_\perp \sim Q \lambda$, and the $\bar n$-collinear field
contains massive gauge bosons 
moving near the $\bar n$-direction, with momentum scaling $n \cdot p \sim Q$,
$\bar n \cdot p \sim Q \lambda^2$, $p_\perp \sim Q \lambda$. Here we have
$\lambda\sim M/Q$,  where $\lambda \ll 1$ is the power counting parameter used for the
EFT expansion The mass-mode 
field contains massive gauge bosons with all momentum components scaling as 
$Q \lambda\sim M$. The effective theory discussed here is \scetew\ studied in Refs.~\cite{CGKM1,CGKM2,CKM}, and is similar to $\text{SCET}_{\text{I}}$, but with weak-scale mass modes intead of the ultrasoft modes familiar from QCD. If we would consider the broken $SU(2)$ together with QCD, the effective theory would have additional $n$- and $\bar n$-collinear massless gluons and ultrasoft massless gluons, as in $\text{SCET}_{\text{I}}$. The $n$- and $\bar
n$-collinear massless gluons fields would have the momentum scaling of the
$n$- and $\bar n$-collinear massive gauge fields of the broken  $SU(2)$. The
ultrasoft gluon fields would have the momentum scaling $p^\mu\sim Q\lambda^2$
with $p^2\sim M^4/Q^2$. At $\mu=M$ the $n$- and $\bar n$-collinear massive
gauge fields and the mass-modes can be integrated out, leaving a common
massless $\text{SCET}_{\text{I}}$ theory for $\mu<M$. Such a situation
is realized in the $SU(3) \times SU(2)\times U(1)$ electroweak theory~\cite{CGKM1,CGKM2}. 

The interactions of the mass-mode fields with the collinear fields are described
by mass-mode $S$-Wilson lines whose definition is identical to the $Y$-Wilson
lines that arise for massless ultrasoft modes in massless
$\text{SCET}_{\text{I}}$ upon the ultrasoft field redefinition. The difference
is that the mass-mode Wilson lines contain mass-mode gauge fields rather than
ultrasoft massless gauge fields. Thus the effective field theory current for
the broken $SU(2)$ has the form\footnote{
In the presence of additional ultrasoft massless gauge fields the effective
theory current would have the form  
$J(\omega,\bar \omega,\mu)
= [\bar \xi_{n,\omega} W_{n} S_n^\dagger Y_n^\dagger \Gamma 
  Y_{\bar n}  S_{\bar n}W_{\bar n}^\dagger \xi_{\bar n,\bar\omega}](0)$, with ultrasoft
  $Y$-Wilson lines.
}
\begin{eqnarray}
\label{currentscet}
 J(\omega,\bar \omega,\mu)
   = [\bar \xi_{n,\omega} W_{n} S_n^\dagger \Gamma
  S_{\bar n}W_{\bar n}^\dagger \xi_{\bar n,\bar\omega}](0) \,,
\end{eqnarray}\
where
\begin{eqnarray} 
 S_n^\dagger(x) & =&   {\rm P} \, 
   \exp\Bigl[ i g\int_{0}^\infty ds\, n \cdot A_{s}(ns+x) \Bigr]
\,,\nonumber \\ 
 S_{\bar n}(x)  &=&  \overline {\rm P} \:
   \exp\Bigl[ -i g \int_{0}^\infty ds\, \bar n \cdot A_{s}(\bar n s+x) \Bigr] 
    \,.
\label{Sn}
\end{eqnarray}
More details can be found in Ref.~\cite{Fleming:2007xt}.

\section{Factorization and Collinear Wilson Lines}
\label{sec:wilson}

\begin{figure}
\begin{center}
\includegraphics[width=3.2cm]{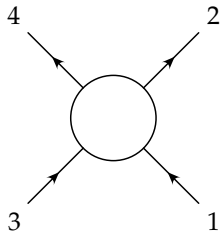}
\end{center}
\vspace{-0.5cm}
\caption{A scattering amplitude with four external particles defining four
  collinear directions $n_1$, $n_2$, $n_3$, and $n_4$. \label{fig:scatter}} 
\end{figure}

Consider a high energy scattering process with two or more particles, in the
$n_i$ direction, $i=1,\ldots ,r$ (see Fig.~\ref{fig:scatter}). 
$n_i$-collinear gauge bosons, which have momentum parallel to particle~$i$ can
interact with particle~$i$, or with the other particles $j\not=i$. The
coupling of $n_i$-collinear gauge bosons to particle~$i$ is included
explicitly in the SCET Lagrangian. The particle-gauge interactions are
identical to those in the full theory, and there is no simplification on
making the transition to SCET. However, if an $n_i$-collinear gauge boson
interacts with a particle $j$ not in the $n_i$-direction, then
particle~$j$ becomes off-shell by an amount of order $Q$, and the intermediate
particle~$j$ propagators can be integrated out, giving a Wilson line
interaction in SCET. The form of these operators was derived in
Ref.~\cite{SCET,SCET2}, and gives the Wilson line interaction $W_{n_i}^\dagger
\xi_{n_i}$, where $W_{n_i}$ is a Wilson line in the $\bn_i$ direction in the
same representation as $\xi_{n_i}$. This is easy to see in processes with only
two collinear particles. But even in complicated scattering processes with
more than two collinear particles the Wilson line interaction still has
the form $W_{n_i}^\dagger \xi_{n_i}$. Gluon emission from all particles other
than the $n_i$-collinear particle combine (using the fact that the operator is
a gauge singlet) to give a Wilson line in the representation of the
$n_i$-collinear particle. 

The Feynman rules for multiple gauge emission of $n_i$-collinear gluons from
particle $j$ gives factors of the form 
\begin{eqnarray}
\frac{\epsilon \cdot n_j}{k \cdot n_j}.
\label{11a}
\end{eqnarray}
The $n_i$-collinear gauge field has momentum $k$ and polarization $\epsilon$
in the $n_i$-direction at leading order in SCET power counting, so the above
expression can be replaced by 
\begin{eqnarray}
\frac{\epsilon \cdot n_j}{k \cdot n_j} &\to & \frac{n_i \cdot n_j}{n_i \cdot n_j} 
\frac{\epsilon \cdot \bar n_i}{k \cdot \bar n_i} =\frac{\epsilon \cdot \bar n_i}{k \cdot \bar n_i} 
\label{11}
\end{eqnarray}
using the leading (first) term in Eq.~(\ref{3}) for the decomposition of both
$k$ and $\epsilon$. This expression is independent of $n_j$. As a result, when
one combines emission from all particles which are not in the $n_i$-direction,
the term in Eq.~(\ref{11}) can be factored out, and the color matrices
combined to form a single Wilson line in the $\bn_i$ direction. This is the
basis for factorization in SCET, since $n_i$-collinear interactions are
independent of the dynamics of all particles not in the $n_i$-direction. 

Unfortunately, Eq.~(\ref{11}), while valid at tree-level, can not be used for
loop diagrams. The reason is that loop diagrams require a regulator for the
Wilson lines.  For example,  with analytic regularization, Eq.~(\ref{11})
becomes 
\begin{eqnarray}
\frac{\epsilon \cdot n_j}{\left(k \cdot n_j\right)^{1+\delta}} &\to &
\frac{n_i \cdot n_j}{\left(n_i \cdot n_j\right)^{1+\delta}}  
\frac{\epsilon \cdot \bar n_i}{\left(k \cdot \bar n_i\right)^{1+\delta}} \nn
&=& \frac{1}{\left(n_i \cdot n_j\right)^{\delta}} 
\frac{\epsilon \cdot \bar n_i}{\left(k \cdot \bar n_i\right)^{1+\delta}}
\label{reg}
\end{eqnarray}
and the $j$ dependence no longer cancels. Thus the identities which allowed
one to combine all the $n_i$-collinear emissions into a single Wilson line in
the $\bar n_i$ direction no longer hold.  

In this paper, we regulate the Wilson lines using the $\Delta$-regulator,
which also introduces $i$-dependence into Eq.~(\ref{11}), and naively breaks
factorization. We will see that after zero-bin subtraction, the $j$-dependence
cancels, and factorization is restored. 

\section{$\Delta$ Regulator}\label{sec:delta}

The $\Delta$-regulator for particle~$i$ is given by replacing the propagator denominators by
\begin{eqnarray}
\frac{1}{(p_i+k)^2-m_i^2} \to \frac{1}{(p_i+k)^2-m_i^2-\Delta_i}.
\label{deltareg}
\end{eqnarray}
This regulator can be implemented at the level of the Lagrangian, since it
corresponds to  a shift in the particle mass. The on-shell condition remains
$p_i^2=m_i^2$. 

In SCET, the collinear propagator denominators have the replacement of
Eq.~(\ref{deltareg}). If particle~$j$ interacts with an $n_i$-collinear gluon
and becomes off-shell, then 
\begin{eqnarray}
\displaystyle
\frac{1}{(p_j+k)^2-m_j^2-\Delta_j} 
\mbox{\hspace{4cm}}
\nonumber\\
\qquad\qquad
\displaystyle \to \frac{1}{\frac12(\bn_i \cdot k) (\bn_j
  \cdot p_j)(n_i \cdot n_j) - \Delta_j}
\,,
\label{a7}
\end{eqnarray}
where $k$ is $n$-collinear, and Eq.~(\ref{11}) becomes
\begin{eqnarray}
\frac{\epsilon \cdot n_j}{k \cdot n_j} &\to &
\frac{\epsilon \cdot \bar n_i}{k \cdot \bar n_i - \delta_{j,n_i}}\,,\nn
\delta_{j,n_i} &\equiv& \frac{2\Delta_j}{(n_i \cdot n_j)(\bn_j \cdot p_j) }\fs
\label{eq7}
\end{eqnarray}
The denominator in Eq.~(\ref{11a}) gets shifted by $\delta_{j,n_i}$, as can be
seen from the denominator of Eq.~(\ref{a7}). The Wilson lines in the
$\Delta$-regulator method will be regulated using Eq.~(\ref{eq7}). As a
result, in the multiparticle case, it is not possible to combine
$n_i$-collinear gluon emission off the various particles into a single Wilson
line in the $\bn_i$ direction, since $\delta_{j,n_i}$ depends on the particle
$j$. However, we will see that after zero-bin subtraction, the $j$-dependence
drops out, and $n_i$-collinear gluon emission can be combined into a single
Wilson line. 

While $\delta_{i,n}$ and $\Delta_i$ are related by Eq.~(\ref{eq7}), it is
useful to retain both variables during the computation. 

\section{Calculation in the effective theory}\label{sec:calc}

\begin{figure*}
\begin{center}
\includegraphics[width=4cm]{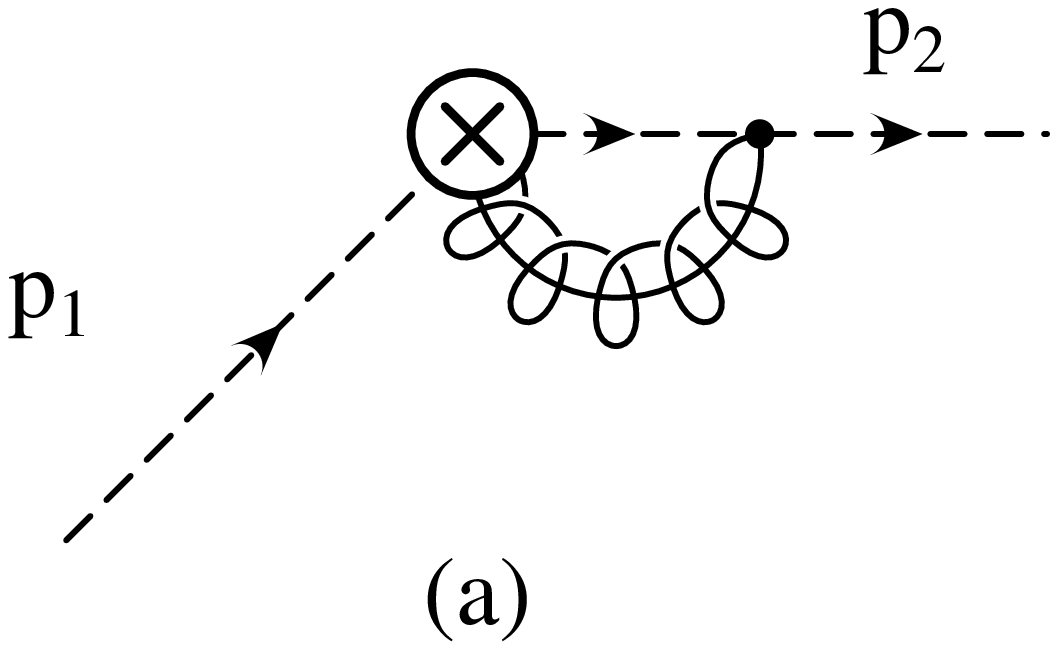}
\qquad\includegraphics[width=4cm]{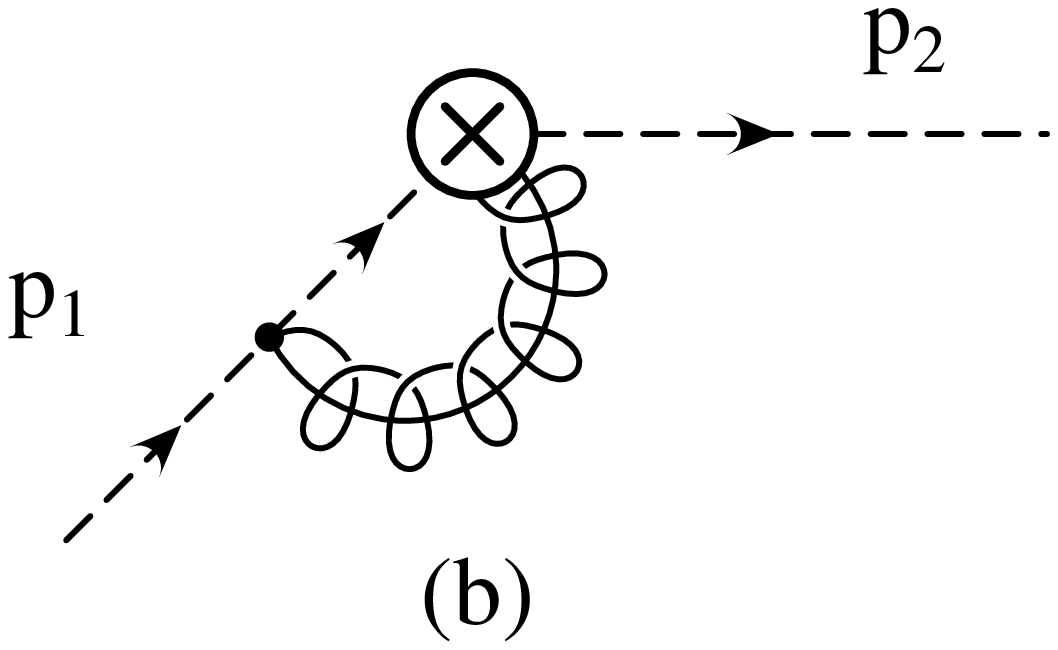}
\qquad\includegraphics[width=4cm]{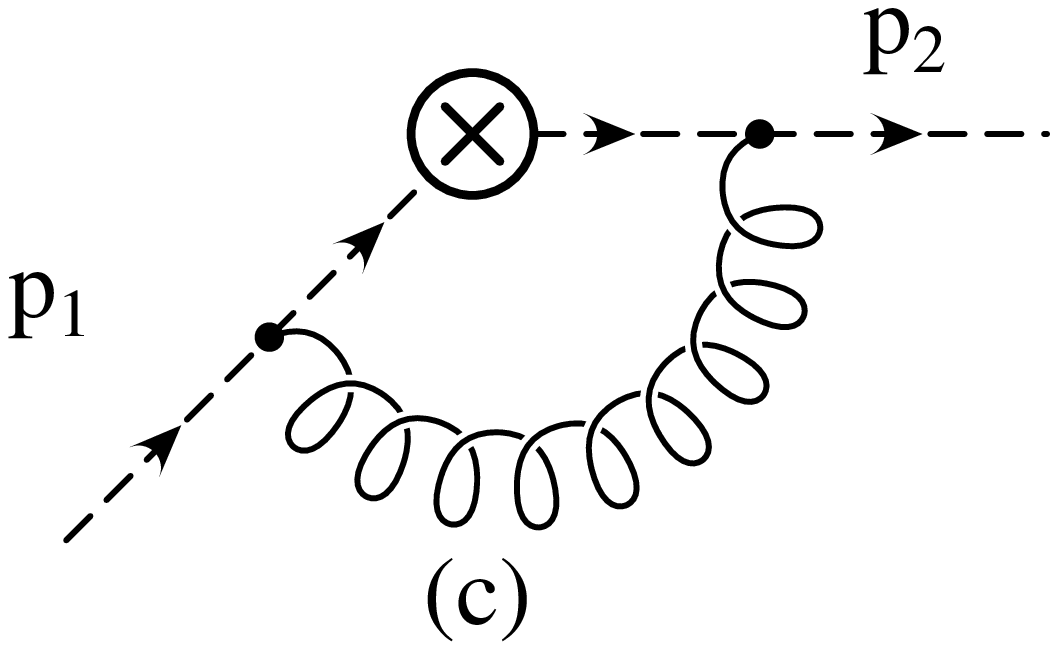}
\end{center}
\vspace{-0.5cm}
\caption{Diagrams of order $O(\al)$ in the effective theory. Wavefunction
  diagrams are not shown. The dashed line denotes a fermion, the spring denotes
  a mass-mode gauge boson and the spring with a line denotes a collinear
  gauge boson.}\label{fig:scet} 
\end{figure*}

The one-loop effective theory vertex graphs are the $n$-collinear,
$\bn$-collinear and mass-mode graphs, shown in Fig.~\ref{fig:scet}. 
The $n$-collinear diagram reads
\begin{eqnarray}
I_n  = -2 i g^2C_F f_\epsilon 
\int { {\rm d}^d k \over (2 \pi )^d} { 1 \over
    \left[ - \bar n \cdot k \right]}
  {\bar n \cdot (p_2-k)  \over\left[ (p_2 - k )^2 \right]}
  {1 \over k^2-M^2} \co
\nonumber\\[-2mm]
\label{1}
\end{eqnarray} 
with $f_\epsilon = \left(4\pi \right)^{-\epsilon}\mu^{2\epsilon}
e^{\epsilon\gamma_E}$. Since the gauge boson is massive, this integral is
divergent even if $p_2$ is off-shell in 
$d=4-2\epsilon$ dimensions. This can be seen as follows: Integrate over $k^+$
by contours and perform the substitution $k^- = z 
p_2^-$. Because the poles of $k^+$ lie in the same half-plane for $k^- >
p_2^-$ and $k^- < 0$, one obtains (keeping $p_2^+ \not=0$ to regulate the
integral) 
\begin{eqnarray}
&& I_{n} =  -2a \mu^{2\epsilon} e^{\epsilon\gamma_E}
\Gamma(\epsilon)\nn
&&\displaystyle\times
\int_0^{1} {dz } \frac{1-z}{z}  \left[M^2(1-z)-p_2^2
  z(1-z)\right]^{-\epsilon}
\label{6}
\end{eqnarray}
where $a = C_F \al/(4\pi)$.
For $z \to 0$ this integral diverges as long as the gauge boson is massive $M
\not =0$, even if $p_2^2 \not=0$. For massless gauge bosons, $M = 0$, and the
factor $z^{-\epsilon}$ from the $p_2^2\not=0$ term regulates the integral when
the fermion is off-shell. 

Introducing the $\Delta$-regulator, the $n$-collinear diagram becomes
\begin{eqnarray}
&& I_n  = -2 i g^2C_F f_\epsilon\nn
&&\times \int { {\rm d}^d k \over (2 \pi )^d} { 1 \over
    \left[ - \bar n \cdot k -\delta_1\right]}
  {\bar n \cdot (p_2-k)  \over\left[ (p_2 - k )^2-\Delta_2 \right]}
  {1 \over k^2-M^2} \fs\nn
\label{1reg}
\end{eqnarray} 
Doing the integrations in exactly the same way as described above, one obtains
for the $n$-collinear integral with an on-shell external fermion $p_2^2=0$
\begin{eqnarray}
I_{n}
&=&  -2a \Gamma(\epsilon) (M^2)^{-\epsilon} \mu^{2\epsilon} e^{\epsilon\gamma_E}
\int_0^{1} {dz } \ {  (1-z)^{1-\epsilon}
\over  z+\delta_1/p_2^-}\nn 
&=&  a \left[\left(\frac{2}{\epsilon}-2\lM\right)\left(1+\log
  (\delta_1/p_2^-)\right)-\frac{\pi^2}{3}+2 \right] \fs \nn
\label{eq:In}
\end{eqnarray}
Note that $\Delta_{1,2}$ and $\delta_1\equiv \delta_{1,n}$, $\delta_2\equiv
\delta_{2,\bn}$ are regulator parameters, and are set to zero unless they are
needed to regulate any divergence.  The regulator parameters are defined using
Eq.~(\ref{eq7}),  
\begin{eqnarray}
\delta_1 &\equiv & \delta_{1,n}=2 \frac{\Delta_1}{(n \cdot \bn)(n \cdot
  p_1)}=\frac{\Delta_1}{p_1^+}\co\nn 
\delta_2 &\equiv & \delta_{2,\bn}=2 \frac{\Delta_2}{(n \cdot \bn)(\bn \cdot
  p_2)}=\frac{\Delta_2}{p_2^-}\,.\co 
\label{dreln}
\end{eqnarray}
We recall that
$n_1=\bn$ because $p_1$ is in the $\bn$ direction, and similarly for
$\delta_2$. The terms $\delta_{1}$ and $\delta_2$ transform under boosts as
the $-$ and $+$ component of a vector, and
$\lM, \lQ$ are defined as
\begin{eqnarray}
\lM = \log \frac{M^2}{\mu^2},\qquad
\lQ = \log \frac{Q^2}{\mu^2}.
\end{eqnarray}
Equation~(\ref{1reg}) depends on $\Delta_2$ and $\delta_1$ which regulate the
collinear and Wilson line propagators, respectively. $\Delta_2$ is not needed
to regulate a divergence in the integral, so the result in Eq.~(\ref{eq:In})
only depends on the 
Wilson line regulator $\delta_1$. The $n$-collinear graph depends on the scale
$Q$ via the regulator dependence, 
\begin{eqnarray}
\log \frac{\delta_1}{p_2^-} = \log \frac{ \Delta_1}{p_1^+ p_2^-}=\log \frac{\Delta_1}{Q^2}\fs
\end{eqnarray}

The $n$-collinear particle momentum is $p_2$, but the $n$-collinear graph
Eq.~(\ref{eq:In}) depends on particle~1 via its dependence on the regulator
$\Delta_1$ for particle~1. This leads to a violation of factorization in the
collinear sector, since the $n$-collinear graph depends on the properties of
the other particles. In the multiparticle case, this means that the Wilson
lines for all the other particles cannot be combined into a single Wilson line
in the $\bn$ direction --- the loop contributions from the other particles
each depend on their own regulator $\delta_i$, and the different contributions
cannot be combined into a single amplitude. 

The $n$-collinear wavefunction renormalization graph is identical to that in
the full theory. It does not need any $\Delta$-regularization, and reads 
\begin{eqnarray}
W_n &=& a \left[\frac{1}{\epsilon}-\frac12-\lM \right]\fs
\end{eqnarray}
The normalization convention is such that one gets a contribution of $-W_n/2$
for each external $n$-collinear fermion.

The $\bar{n}$-collinear integral $I_{\bar{n}}$ can be obtained from $I_n$ by
replacing $p_2^-$ by $p_1^+$ and $\delta_1$ by $\delta_2$,
\begin{eqnarray}
I_{\bn}
&=&  a \left[\left(\frac{2}{\epsilon}-2\lM\right)\left(1+\log
  (\delta_2/p_1^+)\right)-\frac{\pi^2}{3}+2 \right] \co \nn
\label{eq:Ibn}
\end{eqnarray}
and the $\bn$-collinear wavefunction renormalization is $W_{\bn}=W_n$.

For the mass-mode diagram, one finds
\begin{eqnarray}
I_s &=&-2i g^2 C_F f_\epsilon\! \int { {\rm d}^d k \over (2 \pi )^d}
{1 \over k^2-M^2}    {1  \over -n\cdot k-\delta_2}  { 1 \over -\bar n \cdot k -\delta_1} \nn
&=& a 
\Biggl[- \frac{2}{\epsilon^2}
+\frac{2}{\epsilon}\log\frac{\delta_1  \delta_2}{\mu^2}+ \lM^2-2\lM\log\frac{
  \delta_1  \delta_2}{\mu^2}+\frac{\pi^2}{6} 
\Biggr] \fs \nn
\label{8b}
\end{eqnarray} 
Again, we only keep the leading terms in $\delta_i$, and the integral depends
on both $\delta_1$ and $\delta_2$. The mass-mode wavefunction contribution
vanishes, $W_s=0$, since $n^2=\bn^2=0$. 

\subsection{Zero-Bin Subtractions}\label{sec:sub}

In the effective theory, the gauge boson fields of the full theory are split up
into several different fields which fluctuate over different scales. In order
to avoid double counting of the mass-modes, one has to subtract the
contribution from the collinear fields with vanishing label momenta
\cite{MS}. The zero-bin subtraction for Eq.~(\ref{1}), which amounts to taking
the soft limit in the integrand of the collinear integral, is 
\begin{eqnarray}
I_{n\o} &=& -2 i g^2C_F f_\epsilon \int { {\rm d}^d k \over (2 \pi )^d} { 1
  \over  \left[ - \bar n \cdot k-\delta_1\right]}\nn 
&&  {1 \over\left[ -n \cdot k-\Delta_2/p_2^-\right]}
  {1 \over k^2-M^2} \co
\label{1z}
\end{eqnarray} 
which is the same as the integral Eq.~(\ref{8b}), with $\delta_2 \to
\Delta_2/p_2^-$. One needs to retain $\Delta_2$ to regulate the singularity,
since the $k^2$ term in the collinear propagator has been expanded out.
Subtracting this from the collinear integral yields 
\begin{eqnarray}
I_n-I_{n\o} &=&a \biggl[
\frac{2}{\epsilon^2}-\frac{2}{\epsilon}
\log\frac{\Delta_2}{\mu^2}+\frac{2}{\epsilon}-2\left(1-\log\frac{\Delta_2}{\mu^2}\right)\lM\nn    
&&- \lM^2 -\frac{\pi^2}{2}+2\biggr] \fs
\label{1sub}
\end{eqnarray}
This combination only depends on the gauge boson mass and the regulator of the
collinear fermion, $\Delta_2$. The zero-bin subtraction $W_{n\o}$ for the
wavefunction renormalization $W_n$ vanishes. 

There are two very important differences between the zero-bin subtracted result
$I_n-I_{n\o}$ and the unsubtracted result $I_n$: The zero-bin subtracted
integral no longer depends on the hard scale $Q$, and it depends only on the
regulator $\Delta_2$ for the $n$-collinear particle,  rather than on the
regulator $\Delta_1$ for the other particle. This means it depends on the
regulator of the collinear particle rather than the regulator of the Wilson
line, and it
implies that zero-bin subtraction restores factorization in the effective
theory. The hard scale has been factored out of the collinear contribution. In
addition, in the multiparticle case, since the collinear graphs only depend on
the $n$-collinear particle regulator ($\Delta_2$ in our problem), the Wilson
line contributions from all the other particles can be combined into a single
Wilson line in the $\bn$ direction, as was naively true at tree-level. This is
because the zero-bin subtracted collinear graph does not need a regulator for
the Wilson line. 

The final result of the effective theory vertex computation  is
\begin{eqnarray}
&&\left(I_n-I_{n\o}\right)+\left(I_{\bar n}-I_{\bar n\o}\right)+I_s\nn
&&-\frac12(W_{n}-W_{n\o})
- \frac12(W_{\bn}-W_{\bn\o})-W_{s} \nn
&=&a\biggl[ \frac{2}{\epsilon^2}+\frac{2}{\epsilon}\log\frac{\delta_1 \delta_2
  \mu^2}{\Delta_1\Delta_2}+\frac{3}{\epsilon}- \lM^2\nn 
&&-2\lM\log\frac{ \delta_1
      \delta_2 \mu^2}{\Delta_1\Delta_2}-3\lM-\frac{5\pi^2}{6}+\frac92\biggr]\co
\label{eq:res}
\end{eqnarray}
where we have added the zero-bin subtracted collinear graphs and the mass-mode
graph. This result has to be independent of the regulators. Indeed, using
Eq.~(\ref{dreln}), Eq.~(\ref{eq:res}) can be simplified to 
\begin{eqnarray}
a\biggl[ \frac{2}{\epsilon^2}-\frac{2}{\epsilon}\lQ +\frac{3}{\epsilon}-
\lM^2+2\lQ \lM-3\lM-\frac{5\pi^2}{6}+\frac92\biggr]\fs\nn 
\label{eq:resa}
\end{eqnarray}
This is the correct effective theory result, and when combined with the
matching computation at $Q$~\cite{dis} correctly reproduces the known 
full-theory computation of the form-factor. 

Note that without zero-bin subtractions, the effective theory result would be
\begin{eqnarray}
&&I_n+I_{\bar n}+I_s-\frac12 W_{n}- \frac12 W_{\bn}-W_{s} \nn 
&=&a\biggl[- \frac{2}{\epsilon^2}+\frac{2}{\epsilon}\log\frac{\Delta_1^2
  \Delta_2^2}{Q^6\mu^2}+\frac{3}{\epsilon}+ \lM^2\nn 
&&-2\lM\log\frac{\Delta_1^2 \Delta_2^2}{Q^6 \mu^2}-3\lM-\frac{\pi^2}{2}+\frac92\biggr]\fs
\label{eq:wrong}
\end{eqnarray}
This is incorrect, and does not reproduce the full theory result when the
matching condition at $Q$ is included. The $1/\epsilon$ singularities, which
are ultraviolet, do not give the correct anomalous dimension
$\gamma=a(4\lQ-6)$ for the current in SCET. The result is also not independent
of the regulator. Idilbi and Mehen~\cite{IM1,IM2} have previously arrived at
the same conclusions for QCD, where the gauge boson is massless. However, the
necessity of zero-bin subtractions becomes more obvious with a massive gauge
boson, since the effective theory graphs are no longer scaleless due to the
gauge boson mass. One can see that Eq.~(\ref{eq:wrong}) also does not give the
correct finite parts of the diagram. 

Using the $\Delta$-regulator, we have seen that the vertex corrections are
$\left(I_n-I_{n\o}\right)+\left(I_{\bar n}-I_{\bar n\o}\right)+I_s$ after
including zero-bin subtractions. Also, $I_{n\o}=I_{\bn\o}=I_s$, so the vertex
correction can be written as $I_n+I_{\bn}-I_s$. Recently, Idilbi and
Mehen~\cite{IM1,IM2} showed in their study of deep inelastic scattering in
QCD, that the combination  $I_n+I_{\bn}-I_s$ does not need any additional
regulator beyond dimensional regularization. The same result continues to hold
for broken $SU(2)$ with massive gauge bosons, where the role of the 
ultrasoft contribution  is adopted by the mass-mode contribution 
$I_s$. Thus the integrand obtained by combining $I_n+I_{\bn}-I_s$ does not
need any $\Delta$-regulator either, and can be evaluated explicitly to give  
\bea\label{eq:r1}
I_n+I_{\bar{n}}-I_s &=&  a \Bigg[ \frac{2}{\epsilon^2}-
  \frac{2}{\epsilon}\lQ + \frac{4}{\epsilon}- \lM^2 +2\lQ \lM \nn 
&&-4\lM -\frac{5\pi^2}{6} +4 \Bigg]
\,.
\eea
When combined with the wavefunction graphs this gives the correct amplitude
Eq.~(\ref{eq:resa}). The details of the computation are given in
Appendix~\ref{app:im}. The form of the expression, $I_n+I_{\bn}-I_s$ in broken
$SU(2)$ -- and similarly in QCD -- is
counter-intuitive if one is used to thinking about effective field theories
using the method of regions. This is because one has to subtract the
mass-mode/ultrasoft region 
from the sum of the collinear regions to get the correct amplitude.

The above discussion shows that in practice one can identify the
respective zero-bin contribution of the collinear integrals $I_{n\o}$ and $I_{\bar{n}\o}$
with the mass-mode integral $I_s$. Doing this identification also in the case
where the integrals are done 
separately with the regulator, one does not need any relations between
$\Delta_i$ and $\delta_i$. Instead one regulates the collinear Wilson lines
and propagators in the soft graphs with $\delta_i$, and 
Eq.~(\ref{1sub}) is instead given by the same expression with $\delta_2 p_2^-$
in place of $\Delta_2$.

\subsection{Momentum Regions}

\begin{figure}
\begin{center}
\includegraphics[width=7cm,bb=40 221 576 749]{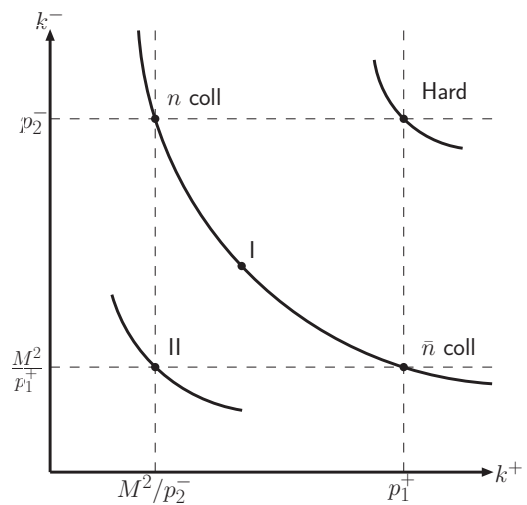}
\end{center}
\caption{Modes which contribute to the effective theory. 
Mass-modes are shown as I.  Additional
 massless gluons contribute modes shown as II.
  }\label{fig:momregions} 
\end{figure}

In this section we discuss the various momentum regions that contribute to the
computation of the Sudakov form factor paying particular attention to the
role of the zero-bin subtractions to the contributions from the $n$- and $\bar
n$ collinear regions. The momentum regions which contribute to the Sudakov
form factor are illustrated in 
Fig.~\ref{fig:momregions}. The hard contribution is the high-scale matching at
$Q$, and the remaining contributions are given by the effective theory. The
effective theory contributions are located
 along the curve $k^2=M^2$. The $n$-collinear contribution arises from $k^-
 \sim p_2^- \sim Q$, so that $k^+ 
\sim M^2/Q$ and the $\bn$-collinear contribution arises from $k^+ \sim p_1^+
\sim Q$, $k^- \sim M^2/Q$. The ultrasoft region with $k^+ \sim M^2/p_2^-\sim
M^2/Q$ and $k^- \sim M^2/p_1^+ \sim M^2/Q$ is not on the $k^2=M^2$ hyperbola,
and does not contribute to the amplitude. Interestingly, it now turns out that
the mass-mode region, with all components of $k$ of order $M$, does not
contribute to the amplitude either.\footnote{It is important to include the mass-mode contribution. The mass-mode region does not contribute, because of a cancellation between the collinear modes and mass-modes in the mass-mode region. See below.} As shown in Ref.~\cite{CGKM1}, the
contributions to the amplitude from the mass-mode region vanish if the
analytic regulator of Eq.~(\ref{reg}) is employed, and only the $n$- and $\bar
n$-collinear regions give nonzero results. As we show below, the same is true
also when the $\Delta$-regulator is used. However, in this case, the
argument is more subtle and requires that zero-bin contributions to
the $n$- and $\bar n$-collinear pieces are properly accounted for.

While the analytic regulator does not introduce any new dimensionful
parameters, the $\Delta$-regulator introduces the new dimensionful
scales $\Delta_{1,2}$, and the picture changes because of contributions from
unphysical regulator dependent regions, which are absent in the total
amplitude. This has been noted previously~\cite{dorsten,messenger}. 
\begin{figure}
\begin{center}
\includegraphics[width=7cm,bb=95 228 609 747]{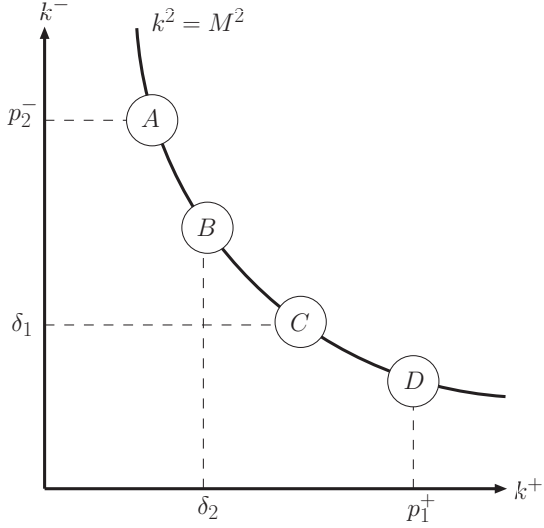}
\end{center}
\caption{Momentum regions which contribute to the effective theory
  integrals. $A$ and $D$ are collinear, and $B$ and $C$ are
  regulator-dependent mass-mode regions.}\label{fig:momregions2} 
\end{figure}
The $n$-collinear graph $I_n$, Eq.~(\ref{eq:In}) gets contributions from $k^-
\sim p_2^-\sim Q$,  $k^+ k^- \sim M^2$ and $k^- \sim \delta_1$, $k^+ k^- \sim
M^2$. This is shown as regions $A$ and $C$ in Fig.~\ref{fig:momregions2}. The
$\bn$-collinear graph gets contributions from regions $D$ and $B$. 

The zero-bin integral $I_{n\o}$ gets contributions from the region with $k^+
\sim \Delta_2/p_2^-$ and from $k^- \sim \delta_1$, i.e.\ from regions $B$ and
$C$. Thus, for the zero-bin subtracted collinear integral $I_n-I_{n\o}$,
region $C$ cancels, and the resulting contributions are from regions $A$ and
$B$. Similarly, $I_{\bn}-I_{\bn\o}$ gets contributions from $D$ and $C$. 
On the other hand, the mass-mode graphs get a contribution from the region
$k^- \sim \delta_1$, $k^+k^- \sim M^2$ and  from 
$k^+ \sim \delta_2$, $k^+ k^- \sim M^2$, i.e.\ regions $B$ and $C$. 
We now see that the total amplitude only gets contributions from $A$ and $D$.
The additional regions $B$ and $C$ introduced by the regulator drop out, as
they should. To achieve the cancellation of the unphysical regions $B$ and $C$
it is essential to account for the zero-bin subtractions for the collinear
regions.

\subsection{Gauge dependence}\label{sec:gauge}

So far, we have been working in Feynman gauge, $\xi = 1$. Let us now analyze
the gauge dependence of the different parts of the effective 
theory calculation by using a general $R_\xi$ gauge with the gauge boson propagator
\beq
i\Delta_{\alpha \beta}(k) = \frac{1}{k^2-M^2}\left[g_{\alpha \beta} +
  \left(\xi-1\right) \frac{k_\alpha k_\beta}{k^2-\xi M^2}\right] \fs
\eeq
In the full theory, the new $\xi$ dependent contribution to the vertex graph,
$I^{(\xi)}$, stemming from the second part of the gauge boson propagator is 
\begin{eqnarray}
I^{(\xi)} 
&=&\frac{\alpha_s}{4\pi} C_F \Gamma J\nn
J &=&-16 \pi^2 i(\xi-1) f_\epsilon \int { {\rm d}^d k \over (2 \pi )^d}
\frac{1}{k^2-M^2}\frac{1}{k^2-\xi M^2}\nn 
&=& \frac{\xi-1}{\eUV}-(\xi-1)\log \frac{M^2}{\mu^2}+(\xi-1)-\xi \log
\xi \fs\nn
\end{eqnarray}
The full theory vertex graph gets shifted, $I_V \to I_V + a J$. There is a
similar shift in the full theory wavefunction contribution, $W \to W+aJ$ so
that the on-shell $S$-matrix element $I_V-W$ is $\xi$-independent. The high
scale matching coefficient is the full theory result with all infrared scales
set to zero, and so is gauge invariant. 

In terms of the method of regions, $J$ only has contributions from the mass
mode region where $k^+\sim k^-\sim k^\perp~\sim M$, given $\xi$ is counted as 
order $O(1)$. Therefore, one might not 
expect this additional piece to show up in the collinear vertex diagrams in
the effective theory. However, doing the calculation of the additional parts
$I_{n}^{(\xi)}$ and $I_{\bar{n}}^{(\xi)}$ of the collinear integrals  yields
(see Ref.~\cite{SCET} for the Feynman rules)
\bea
I_{n}^{(\xi)} &=& -ig^2 C_F f_\epsilon \int \frac{d^dk}{(2\pi)^d}
\left[n^\alpha-\frac{\gamma^\alpha_\perp \slashed{k}_\perp
  }{\bar{n}\cdot(p_2-k)} \right] \nn
&&\times \frac{\bar{n}\cdot(p_2-k)}{(p_2-k)^2-\Delta_2}\Gamma
\frac{1}{-\bar{n}\cdot k-\delta_1} \bar{n}^\beta \frac{1}{k^2-M^2}\nn
&& \times \left[
  \left(\xi-1\right) \frac{k_\alpha k_\beta}{k^2-\xi M^2}\right]\nn
&=& \frac{\al}{4\pi}C_F \Gamma J \co
\label{Inxi}
\eea 
and similarly for $I_{\bar{n}}^{(\xi)}$. Note that for Eq.~(\ref{Inxi}) 
we have adopted $p_2^+=p_2^\perp=0$.

For the mass-mode diagram, the additional piece reads
\begin{eqnarray}
I_{s}^{(\xi)}&=& -ig^2 C_F f_\epsilon \int \frac{d^dk}{(2\pi)^d}
\frac{1}{-n\cdot k-\delta_2} \Gamma  \frac{1}{-\bar{n}\cdot k -
  \delta_1} \nn
&&\times \frac{1}{k^2 -M^2} \left[(\xi-1) \frac{(n\cdot k)( \bar{n} \cdot k)}{k^2 -
  \xi M^2} \right]\nn
&=& \frac{\al}{4\pi} C_F \Gamma J \fs
\end{eqnarray}

The collinear and soft wavefunction renormalization graphs also get shifted by
$J$, $W_n \to W_n + a J$, $W_{\bn}\to W_{\bn}+aJ$, $W_s \to W_s + a J$. 

Without accounting for zero-bin subtractions, the effective theory result 
$I_n+I_{\bn}+I_s-(W_n/2+W_{\bn}/2+W_s)$ would get shifted by
$aJ+aJ+aJ-(aJ/2+aJ/2+aJ)=aJ$, and is not gauge invariant. With zero-bin
subtractions, however, the effective theory collinear vertex graph is
$I_n-I_{n\o}$, 
which is gauge invariant, since both terms shift by $aJ$. Similarly, the
collinear wavefunction graph is $W_n-W_{n\o}=W_n-W_s$ which is also gauge
invariant, since again both terms are shifted by $aJ$. Thus, the collinear
vertex and wavefunction contributions are each separately gauge invariant. The
mass-mode contributions $I_s$ and $W_s$ are each shifted by $aJ$, so the net
soft contribution $I_s-W_s$ is gauge invariant as well. 

Thus zero-bin subtractions are also necessary to maintain gauge invariance of 
the two collinear and the mass-mode sectors of the effective theory, as
required by factorization.

\section{Conclusions}\label{sec:sum}

SCET with massive gauge bosons requires an additional regulator on top of the
common dimensional regularization to obtain well defined expressions for
individual Feynman diagrams. In this work have proposed the
$\Delta$-regulator
to regularize the singularity from the Wilson line propagators. Using the $\Delta$-regulator,
the effective theory only gives the correct result for the scattering
amplitude if zero-bin subtractions for the $n$- and $\bar n$-collinear
contributions are included. For the Sudakov form factor in a broken $SU(2)$
gauge theory with a common gauge boson mass $M$, the total amplitude then is
$I_n+I_{\bar n}-I_s$ as a result of the subtractions. Here $I_n$ and $I_{\bar
  n}$ refer to the $n$- and $\bar n$-collinear diagrams and $I_s$ refers to
the contributions of the mass-mode graphs related to gluons with momenta
$k^+\sim k^-\sim 
k^\perp\sim M$. This result is in analogy to previous results obtained for
unbroken gauge theories with massless gauge bosons~\cite{IM1,IM2}. The result
appears to contradict the method-of-regions approach, where one has to
add up the contributions from all different regions. 

We have demonstrated that one needs to subtract the 
mass-mode region from the sum of the collinear regions to avoid
double-counting, and that gauge invariance of
the effective theory is only maintained if the zero-bin subtractions are
accounted for. 

Zero-bin subtractions also restore factorization between the different
collinear sectors. Naively, one implements factorization by redefining the
collinear fields as~\cite{SCET,SCET2} 
\begin{eqnarray}
W_n^\dagger \xi_n &\to& S_n W_n^{(0)\dagger} \xi_n^{(0)} 
\end{eqnarray}
where the mass-mode fields are in the Wilson line $S_n$, and no
longer couple to the  collinear fields in $W_n^{(0)}$ and $\xi_n^{(0)}$. This
redefinition is not valid at the loop-level, because the regulator dependence
of the collinear graphs breaks factorization. Factorization is restored after
zero-bin subtraction, and thus the proper 
replacement is  
\begin{eqnarray}
\label{replaceWn}
W_n^\dagger \xi_n &\to& S_n \left[ W_n^{(0)\dagger} \xi_n^{(0)}\right]_{\o}
\end{eqnarray}
where the subscript $\o$ is a reminder that the collinear sector requires
zero-bin subtraction.\footnotemark

\bigskip

A.F. was supported by Schweizerischer Nationalfonds.

\begin{appendix}

\section{Calculation without a regulator}\label{app:im}

In this appendix, we calculate the effective field amplitude including
zero-bin subtractions by first adding and then performing the integration. No
regulators are needed in this case, as in the massless case~\cite{IM1,IM2}. 
\begin{widetext}
\begin{eqnarray}
R &=&I_n+I_{\bn}-I_s \nn
&=& -2ig^2 C_F f_\epsilon \int { {\rm d}^d k \over (2 \pi )^d}\Biggl[
   \frac{p_1^+-k^+}{(-p_1^+k^-+k^2)(-k^+)(k^2-M^2)}   +\frac{p_2^--k^-}{(-p_2^-k^++k^2)(-k^-)(k^2-M^2)}
- \frac{1}{(-k^+)(-k^-)(k^2-M^2)}\Biggr]\nn
&=& -2ig^2 C_Ff_\epsilon  \int { {\rm d}^d k \over (2 \pi )^d}\Biggl[
    \frac{2k^2+p_1^+p_2^--p_1^+k^--p_2^-k^+}{(-p_1^+k^-+k^2)(-p_2^-k^++k^2)(k^2-M^2)}
-
\frac{k^4}{(-p_1^+k^-+k^2)(-p_2^-k^++k^2)(-k^+)(-k^-)(k^2-M^2)}\Biggr] \fs\nn
\label{eq:1}
\end{eqnarray} 
\end{widetext}
The total integral is IR finite. It can be decomposed as
\bea\label{eq:r}
R &=& -2ig^2 C_F\left[Q^2 I_0 -2(p_1+p_2)_\mu I^\mu_1 +2 I_2  -I_3\right]\nn
\eea
with 
\begin{eqnarray}
I_0 &=&
f_\epsilon\int { {\rm d}^d k \over (2 \pi )^d} \frac{1}{(k-p_1)^2(k-p_2)^2 (k^2-M^2)}\nn
I_1^\mu &=&
f_\epsilon\int { {\rm d}^d k \over (2 \pi )^d} \frac{k^\mu}{(k-p_1)^2(k-p_2)^2 (k^2-M^2)}\nn
I_2 &=&
f_\epsilon \int { {\rm d}^d k \over (2 \pi )^d} \frac{k^2}{(k-p_1)^2(k-p_2)^2
   (k^2-M^2)}\\
I_3 &=&
f_\epsilon \int { {\rm d}^d k \over (2 \pi )^d}
 \frac{k^2}{(-p_1^+k^-+k^2)(-p_2^-k^++k^2)(-k^+)(-k^-)}\fs \nonumber
\end{eqnarray}
Since we are not interested in subleading terms in $M^2/Q^2$ for all these
integrals, we simplify the last part of the integrand in Eq.~(\ref{eq:1}) and
set $M=0$ to obtain $I_3$. One finds 
\footnotetext{We note that in the presence of massless
  ultrasoft gauge fields the RHS of Eq.~(\ref{replaceWn}) reads
  $Y_n S_n \left[ W_n^{(0)\dagger} \xi_n^{(0)}\right]_{\o}$, with an ultrasoft
  $Y$-Wilson line.}
\begin{eqnarray}
I_0 &=& -\frac{i}{16\pi^2} \int_0^1 {\rm d} z \int_0^1 {\rm d} x 
\frac{z\,}{Q^2 x (1-x)z^2+M^2(1-z)} \nn
&=& -\frac{i}{16\pi^2 Q^2} J_1 \co \nn
I_1^\mu &=&
-\frac{i}{16\pi^2} \frac12\int_0^1 {\rm d} z \int_0^1 {\rm d} x 
\frac{z^2 (p_1+p_2)^\mu}{Q^2 x (1-x)z^2+M^2(1-z)} \nn
&=& -\frac{i}{16\pi^2 Q^2} J_2 \frac12 (p_1+p_2)^\mu \co \nn
I_2 &=& M^2 I_0 +f_\epsilon \int { {\rm d}^d k \over (2 \pi )^d}
\frac{1}{(k-p_1)^2(k-p_2)^2} \nn
&=& M^2 I_0 +  \frac{i}{16\pi^2}
\left[\frac{1}{\epsilon}+\log\frac{\mu^2}{Q^2}+2\right] \co \nn
I_3 &=& -\frac{i}{2} \int_0^{p_2^-} \frac{{\rm d}k^-}{2\pi}\frac{{\rm d}^{d-2}k_\perp}{(2\pi)^{d-2}} \nn
&&\frac{p_1^+ k^-}{[\mathbf{k}_\perp^2+p_1^+k^-][(k^-)^2p_1^+-p_2^-\mathbf{k}_\perp^2-p_1^+p_2^-k^-] }\nn
&& +\frac{i}{2} f_\epsilon \int_{p_2^-}^\infty \frac{{\rm d}k^-}{2\pi}
\frac{{\rm d}^{d-2}k_\perp}{(2\pi)^{d-2}}
\frac{1}{(-p_1^+k^--\mathbf{k}_\perp^2)(k^-)}\nn
&=&  -\frac{i}{16\pi^2} \left[\frac{1}{\epsilon^2} -
  \frac{1}{\epsilon}\log \frac{Q^2}{\mu^2} + \frac12 \log^2
  \frac{Q^2}{\mu^2}-\frac{\pi^2}{12} \right] \co
\end{eqnarray}
with
\begin{eqnarray}
J_n &=& \int_0^1 {\rm d} z \int_0^1 {\rm d} y
\frac{4 z^n}{z^2 (1-y^2) + \lambda^2 (1-z)}
\end{eqnarray}
and $\lambda^2=4M^2/Q^2$. To calculate the integral $J_1$, integrate first
over $y$ and substitute  $w = z + \sqrt{z^2+\lambda^2(1-z)}$, leading to 
\bea
J_1 &=& \frac{\pi^2}{3}+\frac12 \log^2 \frac{Q^2}{M^2} \fs
\eea
For $J_2$, one can simply expand in $\lambda$ after the integration over $y$ to obtain
\begin{eqnarray}
J_2 &=& 2 \log \frac{Q^2}{M^2}-2 \fs
\end{eqnarray}
Adding everything up, one finally finds
\bea
R &=& a \Bigl[\frac{2}{\epsilon^2}-
  \frac{2}{\epsilon}\lQ + \frac{4}{\epsilon}- \lM^2 +2\lQ \lM -4\lM
  -\frac{5\pi^2}{6} +4 \Bigr] \fs\nn
\eea

\end{appendix}

\bibliography{biblio}

\end{document}